# HERITRACE: Tracing Evolution and Bridging Data for Streamlined Curatorial Work in the GLAM Domain

Arcangelo Massari[1,2], Silvio Peroni[1,2]

[1] Digital Humanities Advanced Research Centre(/DH.arc), Department of Classical Philology and Italian Studies, University of Bologna, Bologna, Italy - arcangelo.massari@unibo.it

[2] Research Centre for Open Scholarly Metadata, Department of Classical Philology and Italian Studies, University of Bologna, Bologna, Italy - silvio.peroni@unibo.it

**ABSTRACT**
HERITRACE is a semantic data management system tailored for the GLAM sector. It is engineered to streamline data curation for non-technical users while also offering an efficient administrative interface for technical staff. The paper compares HERITRACE with other established platforms such as OmekaS, Semantic MediaWiki, Research Space, and CLEF, emphasizing its advantages in user friendliness, provenance management, change tracking, customization capabilities, and data integration. The system leverages SHACL for data modeling and employs the OpenCitations Data Model (OCDM) for provenance and change tracking. Future developments include the integration of a robust authentication system and the expansion of data compatibility via the RDF Mapping Language (RML), enhancing HERITRACE's utility in digital heritage management.

**KEYWORDS**
Data Management System - Data Curation - Provenance - Change Tracking - Semantic Web Technologies

## 1. INTRODUCTION

In this paper, we introduce HERITRACE (Heritage Enhanced Repository Interface for Tracing, Research, Archival Curation, and Engagement), a novel semantic data management system which addresses the increasing complexities faced by cultural heritage institutions including galleries, libraries, archives, and museums (GLAM). This system has been developed to support the digital landscape of curating metadata in GLAM institutions. Traditionally, GLAM experts have relied on their interpretative skills and domain knowledge to curate metadata. However, the digitization of cultural heritage data has introduced new challenges, including the representation of data in various machine-readable formats and their preservation in heterogeneous databases. This scenario has created a barrier for domain experts without computer knowledge and, in particular, expertise in Semantic Web technologies. Indeed, these technologies, despite their potential, are complex and have resulted in a paradoxical situation. On one hand, these technologies have made human intervention more critical due to the semantic interpretation of data that cannot be automated. On the other hand, they have limited the number of curators to those who are experts in the Semantic Web.

This technological advancement has led to two contrasting scenarios in the GLAM sector. Some collections have adopted Semantic Web technologies, requiring more staff with technical expertise for long-term maintenance. Others have refrained from adopting these technologies to avoid curatorial complexities. Examples of these two scenarios have been addressed in the FICLIT Digital Library [1] and OpenCitations [2], two infrastructures handled by the University of Bologna. The FICLIT Digital Library, managed via Omeka S, faces limitations due to its simplistic semantic tools and lack of SPARQL query capabilities, leading to challenges in change tracking and transparent provenance management. In contrast, OpenCitations fully embraces Semantic Web technologies but grapples with the issue of incorrect or missing data, a problem that requires human discernment for correction.

The central problem that arises from these scenarios is the gap between the complex digital technologies and the domain expertise of GLAM professionals. This gap hinders effective data curation and limits the potential of digital collections to represent and disseminate cultural heritage accurately and comprehensively. The solution proposed in this project is the development of a framework that facilitates domain experts without skills in Semantic Web technologies in enriching and editing such semantic data intuitively, irrespective of the underlying ontology model and the technologies adopted for storing such data.

The challenges are manifold. A critical goal is to create a system that is user-friendly for several kinds of end-users, including librarians, museologists, gallery curators, archivists, administrators and IT professionals who are tasked with setting up and maintaining the framework. Another significant challenge is provenance management. In the context of GLAM institutions, where the historical and source context of data is paramount, a data management system must accurately track and document the responsible agents and primary data sources. Change tracking is also a fundamental requirement. The system needs to efficiently monitor and record all modifications to the data, allowing for transparency and accountability in the curation process. Customization is a further challenge that a data management system for cultural heritage must address. Recognizing that different GLAM domains have unique requirements for how resources are represented and managed, a customizable interface should be tailored to various data models, enabling the representation of diverse resource types according to specific domain needs. Finally, interfacing with pre-existing data presents a substantial challenge, as GLAM institutions often already possess vast collections, which organize their data through the adoption of different data models. This requirement is particularly important for ensuring that the transition to a new data management system is smooth and does not disrupt the ongoing operations of the institution.

## 2. COMPARATIVE ANALYSIS OF SEMANTIC DATA MANAGEMENT SYSTEMS

The subsequent sections of the paper delve into the specifics of how HERITRACE addresses these challenges. The system's design and functionality are detailed, comparing it with other prominent platforms – i.e. OmekaS [3], Semantic MediaWiki [4], Research Space [5], and CLEF [6]– in terms of user-friendliness, provenance management, change tracking, customization, and data interfacing. These evaluation criteria, employed for the comparative analysis of HERITRACE, are based on those used to assess the CLEF system. This ensures that our assessment criteria are not only relevant but also consistently applied across similar platforms within the digital heritage domain, as summarized in Table 1.

| Name | User friendly (Users) | User friendly (Admin) | Provenance Mgmt. | Change-tracking | Customization | Heterogeneous data sources |
|---|---|---|---|---|---|---|
| OmekaS | ✓ | ✓ | | | ✓ | |
| Semantic MediaWiki | ✓ | ✓ | ✓ | ✓ | ✓ | |
| Research Space | ✓ | | ✓ | | ✓ | ✓ |
| CLEF | ✓ | ✓ | ✓ | ✓ | | |
| HERITRACE | ✓ | ✓ | ✓ | ✓ | ✓ | ✓ |

Table 1: Comparison of Data Management System Features for the GLAM Sector

OmekaS, recognized for its user-friendly interface, primarily serves museums and educational institutions with its intuitive web-publishing platform. However, it exhibits certain limitations in more complex operational aspects. Notably, OmekaS does not inherently track provenance. This limitation can affect the credibility and traceability of the information presented. Additionally, OmekaS lacks inbuilt change-tracking capabilities. Data interfacing in OmekaS presents another challenge. To import pre-existing data in bulk, users must rely on the CSV Import plugin [7]. This plugin necessitates restructuring the original data to fit its specific format with mandatory field names, which can be a cumbersome and time-consuming process. This requirement for data formatting reduces the platform's flexibility in handling heterogeneous data sources.

Semantic MediaWiki significantly enhances the popular MediaWiki platform by integrating semantic capabilities. This blend of features balances user-friendliness for non-technical end-users and the more complex needs of technical administrators. One of the key strengths of Semantic MediaWiki is its customization potential, although it requires a degree of familiarity with both the MediaWiki environment and underlying semantic concepts. In terms of data provenance management, Semantic MediaWiki provides robust support. However, its capabilities for change tracking are not native to the system but are instead supplemented through the use of external plugins. A notable example is the Semantic Watchlist plugin [8], which effectively monitors changes within the wiki. These changes are stored in a relational database rather than in RDF format, which, while practical for tracking purposes, may not align seamlessly with the semantic structure of the data. This discrepancy could potentially restrict the depth of change analysis and the ability to contextualize changes within the semantic framework of the data. Addressing the interfacing with heterogeneous data sources, Semantic MediaWiki initially focused solely on importing OWL ontologies. To broaden its RDF support, the RDFIO extension was introduced [9]. This extension enables the loading of RDF triples, but it is confined to the N-Triples format and notably lacks support for named graphs. This limitation is significant as it restricts the platform's adaptability in various environments that may require more complex semantic data structures.

Research Space, tailored for the academic and research community, excels in user-friendliness for end-users, offering diverse data visualization options such as graphs and temporal maps. However, it maintains a level of complexity for administrators, demanding a steep learning curve. The platform requires a solid understanding of HTML, handlebars and other ResearchSpace-specific components for creating templates, which may be cumbersome even for those with technical expertise. In terms of data provenance, Research Space automatically associates data with its source, ensuring traceability and credibility. However, it lacks a change-tracking system, which could limit its effectiveness in environments where monitoring data modifications over time is crucial. Regarding data interfacing, Research Space allows uploading RDF data directly, which is advantageous for projects involving such formats. However, after the data is uploaded, an administrator's intervention is required to customize the interface appropriately to display the items correctly. This aspect indicates that while Research Space can interface with heterogeneous data sources, doing so involves a significant level of programming complexity for system administrators.

CLEF is designed to manage complex digital libraries, archives, and research data, particularly in the humanities. It offers an administrator-friendly interface and focuses on user-friendliness for end-users, making it suitable for a wide range of audiences within its domain. CLEF's provenance management is robust, utilizing named graphs. Moreover, it does feature change tracking capabilities, including synchronization with GitHub, but lacks a direct system to restore previous versions. Expanding on the capabilities of CLEF, it is important to note that this system does not allow for extensive customization. Moreover, unlike some of its counterparts, CLEF is not designed to upload and manage pre-existing RDF data as-is. This limitation is significant because the software is structured to add items one by one from scratch directly through the user interface. This approach, while potentially beneficial for building new databases, limits the platform's ability to seamlessly integrate and manage existing large-scale datasets. Furthermore, even though CLEF does not impose a specific data model, it organizes data in a format akin to nanopublications for managing provenance. This structure means that if a pre-existing triplestore is connected, the system is not readily equipped to explore the data without a prior reorganization to make it compatible with CLEF's framework.

HERITRACE is designed with a focus on usability for domain experts in the fields of archives, libraries, and museums, who may not possess technical skills. Provenance and change management are handled using the OpenCitations Data Model (OCDM) [10], ensuring reliable tracking and documentation of data changes. Lastly, HERITRACE functions seamlessly with RDF data, enabling it to interface with diverse data sources. Additionally, it is user-friendly for administrators responsible for its configuration. The system operates out-of-the-box with RDF data present on any triplestore. For further customization of the user experience, especially regarding data editing forms, HERITRACE utilizes SHACL [11], a well-known language for validating RDF graphs. This approach eliminates the need to learn a specialized language, as with Research Space, while offering a flexibility level that sits between CLEF and Research Space. HERITRACE also allows predefined graphical modifications through YAML [12] configuration files, simplifying the customization process.

In particular, SHACL allows administrators to specify the classes in the data model adopted for describing the data, properties for each class, and constraints for each property. These constraints include the minimum and maximum number of each property for a specific class, the number of permissible values, or the type of values (e.g., class of the value or datatype). Once these SHACL definitions are in place, HERITRACE updates to display editing forms that enforce the constraints defined in the SHACL document.

Furthermore, HERITRACE enables further customization of the interface through a YAML configuration file. This file allows for the definition of user-friendly names for each class and property. It also enables the specification of whether a property should be displayed and, if so, how it should be represented through a SPARQL query. Properties with an inherent order, such as authors, can have their order predicates defined, allowing the interface to present mechanisms for reordering these elements.

HERITRACE automatically manages provenance and change tracking by leveraging the OCDM to ensure meticulous documentation and traceability of data alterations. Each time an entity is created or modified, a new snapshot is generated and stored within a provenance named graph. Classified as `prov:Entity`, these snapshots link to their corresponding entities via the `prov:specializationOf` property. They record essential timestamps, including their creation (`prov:generatedAtTime`) and when they become invalid (`prov:invalidatedAtTime`). The individuals responsible for data changes are documented using the `prov:wasAttributedTo` property, enhancing accountability and transparency. Crucially, the `prov:hasPrimarySource` property is employed to trace back to the primary sources of the data, establishing a clear lineage and source of information. This feature is vital for maintaining a continuous historical evolution of each entity. Snapshots are connected to their preceding versions through the `prov:wasDerivedFrom` property, allowing for a chronological tracking of changes.

Furthermore, the OCDM framework enhances the Provenance Ontology's capabilities by introducing the `oco:hasUpdateQuery` property. This innovation is pivotal in recording changes to an RDF graph, specifically additions and deletions, through SPARQL `INSERT` and `DELETE` queries. This mechanism facilitates the restoration of entities to specific snapshots by reversing operations from all subsequent updates.

HERITRACE's interface also incorporates a timeline feature, as shown in Figure 1, enabling users to explore different versions of the data. This visual representation lets users discern changes between versions at a glance. If a user chooses to restore an entity's previous version, HERITRACE generates a new snapshot. This new snapshot cites the restored snapshot as its primary source, thus maintaining a coherent and traceable record of the data's evolution.

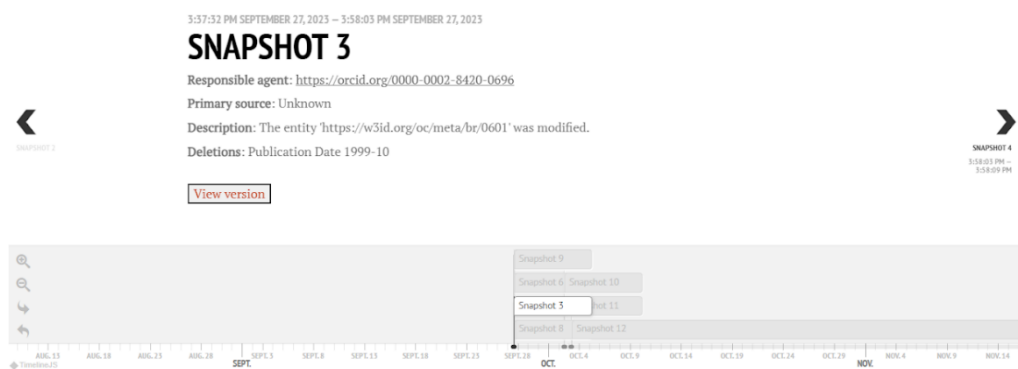

Figure 1: HERITRACE Timeline Interface - This view shows a sequence of data snapshots, allowing users to navigate through versions and view critical metadata for transparent tracking of data provenance and changes.

3.  DISCUSSION AND CONCLUSION

Looking towards the future, HERITRACE is poised for significant enhancements. A crucial area of development is the incorporation of an authentication system. This system is vital for ensuring that only authorized personnel can modify metadata, thus maintaining the integrity and credibility of the information. The proposed solution, RCIAM (Identity Access Management for Research Communities) [13], is set to be adopted by the European Open Science Cloud. It will provide a robust framework for user authentication and authorization, leveraging established protocols like OpenID Connect, SAML, and OAuth [14]. This will enable the organization of users into groups, assignment of roles, and management of access rights, enhancing the security and efficiency of data management.

Another pivotal area of development is the integration of RML (RDF Mapping Language) [15] to extend HERITRACE's capabilities beyond native RDF data. This enhancement aims to broaden the system's adaptability to various data formats, particularly tabular formats like CSV and relational databases. The extension of RML is not just about converting different data formats into RDF; it's also about enabling their modification. This advancement is crucial for projects dealing with a wide range of data types, as it will allow for more flexible and comprehensive data handling.

In addition to the future enhancements already outlined for HERITRACE, an important area for further development is the focus on User Experience Insights. Gaining a deeper understanding of how GLAM professionals interact with HERITRACE can provide invaluable feedback for continuous improvement. This involves actively seeking out and analyzing feedback from those who have tested or used the system in real-world scenarios.

Reflecting on the broader implications of HERITRACE's design and functionalities, it is important to consider how such a system aligns with and supports overarching goals of initiatives like the Cultural Heritage Data Space (CHDS) and the European Collaborative Cloud for Cultural Heritage (ECCCH). The CHDS aims to create a unified, accessible, and secure digital space for European cultural heritage, promoting the sharing and utilization of cultural data across borders. This initiative seeks to enhance the visibility and interoperability of cultural heritage assets, facilitating collaboration among cultural institutions. Similarly, the ECCCH is designed to leverage cloud technologies to foster innovation and collaboration in the cultural heritage sector, providing a platform for sharing resources, tools, and data among cultural institutions and researchers across Europe. Both initiatives underscore the importance of accessibility, interoperability, and collaboration in the digital preservation and dissemination of cultural heritage. While HERITRACE is not directly collaborating with these initiatives, its design principles and functionalities support the shared goals of facilitating access to and collaboration on cultural heritage data, thus contributing to the broader ecosystem of digital cultural heritage management.

In summary, HERITRACE presents itself as a practical solution in the field of semantic data management, with a particular focus on the needs of the GLAM sector. The system provides a user-friendly interface that caters to both non-technical and technical users, alongside features such as provenance management, change tracking, and the ability to customize according to specific needs. Its capability to integrate with existing datasets enhances its practicality. Overall, HERITRACE offers a functional approach to managing digital memory and heritage, potentially contributing to a more comprehensive and accessible understanding of cultural heritage in the digital context. Its design and features position it as a useful tool for professionals in the GLAM sector, aiming to simplify and streamline the management of digital content while respecting the intricacies of cultural heritage data.

For those interested in exploring HERITRACE further, the system along with its documentation are available on GitHub and Software Heritage [16], providing essential resources for implementation and use.

## 4. ACKNOWLEDGEMENTS
This work has been partially funded by Project PE 0000020 CHANGES - CUP B53C22003780006, NRP Mission 4 Component 2 Investment 1.3, Funded by the European Union - NextGenerationEU.